\documentclass[sn-nature,iicol]{sn-jnl}
\usepackage{subfigure}
\usepackage{graphicx}%
\usepackage{multirow}%
\usepackage{amsmath,amssymb,amsfonts}%
\usepackage{amsthm}%
\usepackage{mathrsfs}%
\usepackage[title]{appendix}%
\usepackage{xcolor}%
\usepackage{textcomp}%
\usepackage{manyfoot}%
\usepackage{booktabs}%
\usepackage{pdflscape}
\usepackage{algorithm}%
\usepackage{algorithmicx}%
\usepackage{rotating}
\usepackage{algpseudocode}%
\usepackage{listings}%
\usepackage{hyperref}
\usepackage{bookmark}
\usepackage{lmodern}
\usepackage{booktabs}
\usepackage{tabularx}
\usepackage{array}
\usepackage{amsmath}

\begin{document}

\title[Article Title]{Reducing the Complexity of Matrix Multiplication by Quantum Computing}

\author[1]{Jiaqi Yao}
\author[2]{Tianjian Huang}
\author[3]{Tonghe Zhang}
\author*{Ding Liu*}\email{liuding@tiangong.edu.cn}
\affil{\orgdiv{School of Computer Science and Technology}, \orgname{Tiangong University}, \orgaddress{ \city{Tianjin}, \postcode{300387},  \country{China}}}

\abstract{
Matrix multiplication is a fundamental operation in compute-intensive tasks and a key component of modern quantum acceleration frameworks. Here we present a quantum matrix multiplication algorithm based on quantum kernels (QKMM), achieving an elementary gate complexity of \(O(N^2\log_2N)\), with amplitude encoding overhead explicitly included and without assuming a QRAM oracle. This scaling is asymptotically lower than that of the best-known classical matrix multiplication algorithm \(O(N^{2.371339})\). Building upon QKMM, we establish a family of quantum linear algebra operators, including Quantum Vector Inner Product (V${\scriptstyle 2}$V), Quantum Vector-Matrix Multiplication (V${\scriptstyle 2}$M), QKMM (M${\scriptstyle 2}$M),
Quantum One-to-Many Matrix Multiplication(O${\scriptstyle 2}$M) and Quantum Sequential Matrix Multiplication (SMM), providing a unified framework from vector operations to parallel and sequential matrix transformations. Through noiseless simulations, realistic noise modelling and experiments on a superconducting quantum processor, we systematically characterize the numerical accuracy, resource requirements and hardware execution limits of this operator framework. Furthermore, we integrate SMM into deep neural-network inference, enabling intermediate features to propagate coherently across layers without repeated measurement and re-encoding. These results establish a pathway from quantum circuit-level algorithm design to end-to-end coherent computation, providing a quantum computing framework for matrix-centric compute-intensive applications.
}


\maketitle

\unnumbered
\section{Introduction}

\begin{figure*} 
\centering
\includegraphics[width=1.0\textwidth]{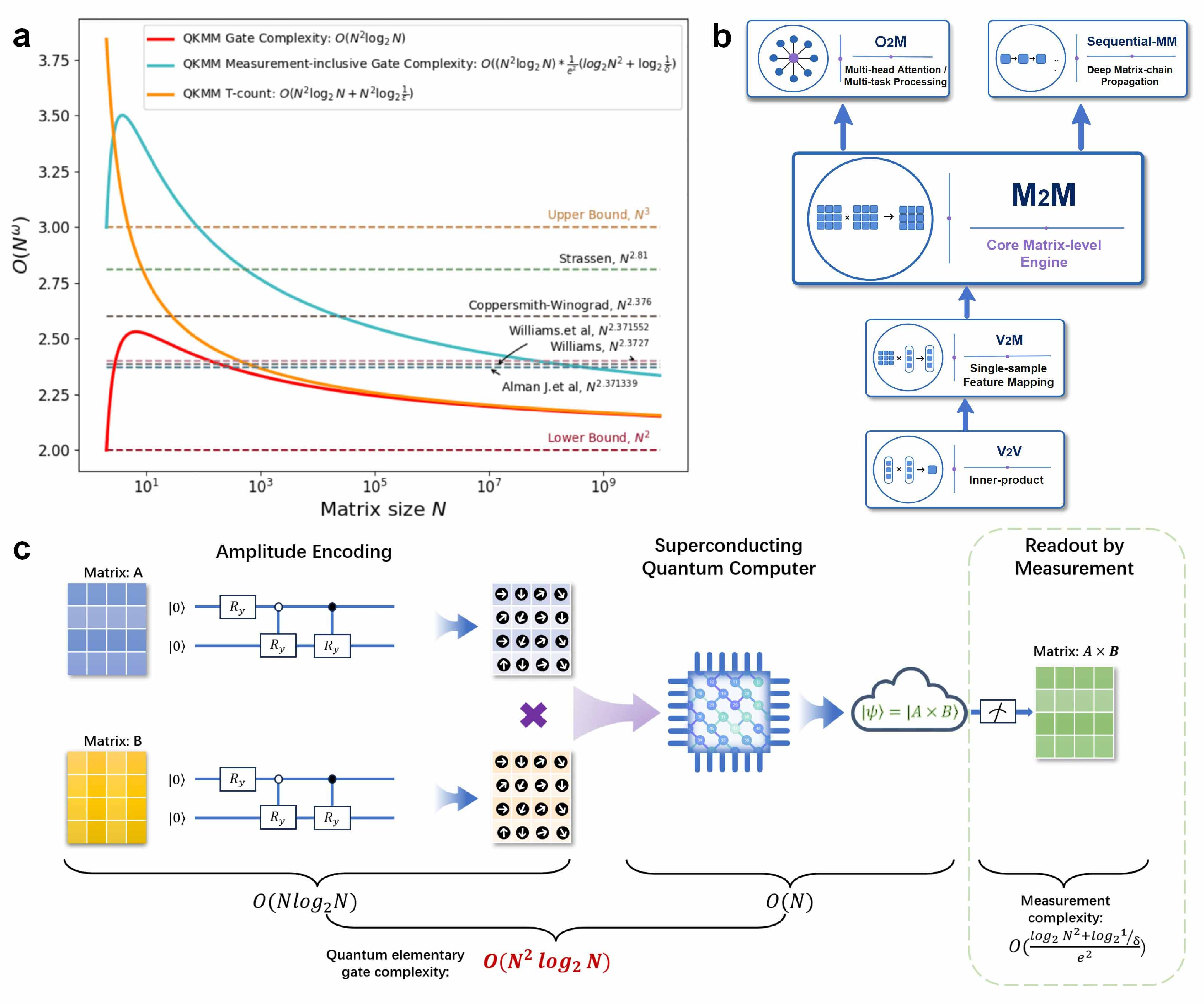} 
	\caption{\textbf{QKMM-centred framework of foundational quantum operators and its computational principles.} \textbf{a}, Comparison of time complexity between the QKMM algorithm and classical matrix multiplication algorithms. \textbf{b}, Hierarchical organization of foundational quantum operators constructed within the QKMM framework. \textbf{c}, Execution pipeline and computational workflow of QKMM.}
	\label{fig:fig1_framework} 
\end{figure*}

For decades, significant efforts have been devoted to improving matrix multiplication efficiency, which has produced a series of landmark achievements \cite{strassen1969gaussian,pan1978strassen,coppersmith1982asymptotic,strassen1987relative,coppersmith1987matrix,williams2012multiplying,le2014powers,alman2024refined,fawzi2022discovering,duan2023faster,williams2024new,alman2025more}. Such research holds particular importance in today's deep learning era, where matrix multiplication is not only a fundamental operation in linear algebra but also the core and computationally intensive component of deep neural networks. Consequently, maximizing the efficiency of matrix multiplication can substantially reduce the computational cost of AI model training. This paper proposes a novel framework that leverages quantum computing to accelerate matrix multiplication. Theoretically, the algorithm achieves an elementary-gate complexity of $O(N^2log_2N)$, yielding near-quadratic scaling and differing from the $O(N^2)$ lower bound for explicit matrix multiplication by only a logarithmic factor. Fig.\ref{fig:fig1_framework}a demonstrates a comparison between this algorithm and numerous historically prominent classical matrix multiplication algorithms.

The pursuit of better matrix multiplication has been marked by a series of groundbreaking achievements since Strassen's revolutionary \( O (N^{2.81} ) \) \cite{strassen1969gaussian}. During the late 20th century, significant advances were made through Pan's trilinear aggregation technique \cite{pan1978strassen} and the Coppersmith-Winograd tensor product approach \cite{coppersmith1982asymptotic}, which reduced the complexity bound to \( O(N^{2.495548}) \). Strassen's subsequent formulation of the relative bilinear complexity theory in 1986 established crucial theoretical foundations for future developments \cite{strassen1987relative}. The 21st century has witnessed remarkable progress through innovative techniques including randomized hashing \cite{williams2012multiplying} and tensor exponent analysis \cite{le2014powers}, culminating in the enhanced laser method, which delivered a time complexity of \( O(N^{2.3728596}) \) \cite{alman2024refined}. A pivotal moment occurred in 2022 when two major breakthroughs emerged simultaneously: DeepMind's AlphaTensor \cite{fawzi2022discovering} leveraged reinforcement learning to discover over 70 novel algorithms, while researchers employed asymmetric hashing to set a new record of  \( O(N^{2.371866}) \) \cite{duan2023faster}. Subsequent work combining recursive laser methods with asymmetric 
hashing further reduced the exponent to 
\( O(N^{2.371552}) \) \cite{williams2024new}. 
More recently, Alman et al. introduced additional asymmetry into 
the laser-method analysis and established the improved bound 
\(O(N^{2.371339})\) \cite{alman2025more}. This five-decade-long pursuit not only has progressively advanced computational complexity boundaries but also has delivered essential technical capabilities for modern compute-intensive domains, particularly deep learning. Complementing these theoretical advances, practical engineering solutions have flourished, with industry leaders developing optimized libraries such as GEMM \cite{vasudevan2017parallel} and QNNPACK \cite{won2022ulppack}, as well as efficient algorithms such as im2col \cite{chellapilla2006high}, to enhance neural network performance.

\begin{figure*} 
\centering
\includegraphics[width=1.0\textwidth]{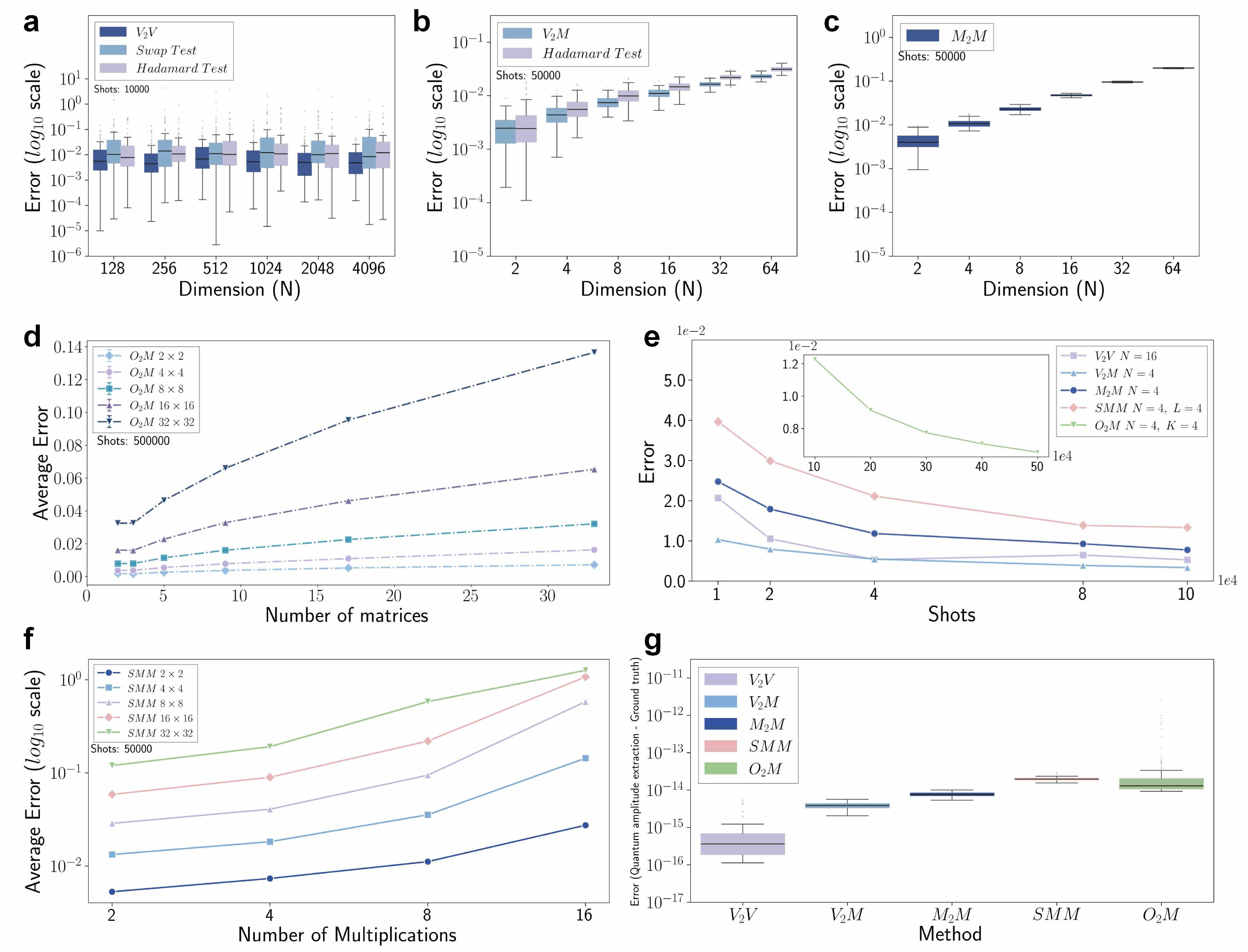} 
	\caption{\textbf{Numerical accuracy and measurement-induced uncertainty of the QKMM operator system under noiseless conditions.} \textbf{a–c}, Relative errors of V{\small 2}V, V{\small 2}M and M{\small 2}M with increasing input dimension $N$, highlighting the relative-error advantages of V{\small 2}V over the Swap Test and Hadamard Test (\textbf{a}) and of V{\small 2}M over the Hadamard Test (\textbf{b}).  \textbf{d}, Cumulative error evolution of O{\small 2}M with increasing number of parallel matrix branches $K$. \textbf{e}, Convergence of relative error with increasing measurement shot counts. f, Cumulative error evolution of SMM with increasing number of sequential matrix multiplications $L$. \textbf{g}, Error distribution of directly extracted quantum amplitudes relative to classical ground truth.}
	\label{fig:fig2_Noiseless} 
\end{figure*}

Evidently, reducing the complexity of classical matrix multiplication has become increasingly challenging. Consequently, researchers have begun to leverage the parallelism of quantum computation to design more efficient matrix multiplication methods, aiming to overcome the current bottleneck. However, the existing research on quantum matrix multiplication (QMM) currently faces a lack of consensus: while classical matrix multiplication complexity is uniformly measured by the number of multiplication operations as a standardized metric, no such established framework exists for evaluating QMM. Various studies adopt divergent approaches, including query complexity (often oracle-based), qubit complexity, gate complexity, and multiplication counting complexity. While early studies \cite{sracic2011quantum} realized a general quantum circuit for matrix multiplication based on modular adders and multipliers, their reliance on circuit stacking incurs a cubic scaling overhead in quantum resources (qubits and gates). This scaling behavior renders such approaches poorly suited for efficient implementation in the current NISQ (Noisy Intermediate-Scale Quantum) era, where hardware resources are severely constrained. In reference \cite{zhang2016quantum}, researchers propose using the hyperentanglement characteristics of optical quantum systems to improve the Swap Test to achieve matrix multiplication, with its complexity measured by time $O(N^2 \epsilon^{-2} \log_{2} \eta^{-1})$, where $\epsilon$ is the statistical error and $\eta$ is the approximation accuracy introduced by Oracle. Hyperentangled states are a promising technology for achieving hyperparallel quantum computation, but they are still in the early stages of development and have only been studied in optical quantum systems and neutral atom systems so far. A recent study utilized the Hadamard Test to design a quantum inner product algorithm, which was subsequently extended to matrix multiplication \cite{xiong2024circuit}. As the time complexity analysis in the article did not explicitly specify the underlying metric units, we conducted a complete analysis of the quantum circuit diagram in this work. Our analysis concludes that the elementary quantum gate complexity in this work is $O(N^3 \log_2 N)$. Table \ref{tab:resource_scaling} summarizes the performance of this work in terms of qubit and elementary quantum gate complexity.
 
We present a Quantum Kernel-based Matrix Multiplication (QKMM) algorithm that reduces the time complexity, measured in quantum gate count, to \(O(N^2\log_2N)\), outperforming the best known classical counterpart, which scales as \(O(N^{2.371339})\) \cite{alman2025more}. 
Centering on the QKMM algorithm, we further construct a family of quantum linear algebra operators, including Quantum Vector Inner Product (V{\small 2}V), Quantum Vector-Matrix Multiplication (V{\small 2}M), QKMM (M{\small 2}M), Quantum One-to-Many Matrix Multiplication (O{\small 2}M) and Quantum Sequential Matrix Multiplication (SMM). This operator family not only lays the foundation for linear algebra and scientific computing, but also covers the vast majority of low-level operations in current deep neural networks.
We conduct comprehensive experiments via noise-free simulation, noisy simulation and physical superconducting quantum computers to analyze the practical performance and error characteristics of QKMM as well as the entire operator suite, which demonstrates the great potential of our proposed algorithm.
Furthermore, we identify an optimal application scenario: deploying SMM for signal propagation in current deep neural networks. Within this scenario, our design circumvents the prohibitive measurement complexity inherent to quantum amplitude encoding, thereby enabling viable quantum speedups.

\begin{table*}[t]
\centering
\caption{\textbf{Resource consumption comparison between foundational quantum operators and baseline architectures.} ($N$ denotes the matrix or vector dimension, $K$ denotes the number of parallel matrices in O{\small 2}M, $L$ denotes the number of chained matrices in SMM, $\varepsilon$ is the synthesis precision, e is the component-wise error, and $\delta$ is the confidence parameter. ``--'' indicates that the corresponding baseline construction is not defined for that operator or that the T-count/Measurement complexity was not evaluated in this comparison.)}
\label{tab:resource_scaling}
\resizebox{\linewidth}{!}{
\begin{tabular}{llccc}
\toprule
\multirow{2}{*}{Operator} 
& \multirow{2}{*}{Metric} 
& \multicolumn{2}{c}{Baseline constructions} 
& \multicolumn{1}{c}{This work} \\
\cmidrule(lr){3-4}\cmidrule(lr){5-5}
& & Swap Test & Hadamard Test & This work \\
\midrule
\multirow{4}{*}{V{\small 2}V} 
& Qubits & $2\log_2 N+1$ & $\log_2 N+1$ & $\log_2 N$ \\
& Gates $\mathcal{O}g$ & $\mathcal{O}(N\log_2 N)$ & $\mathcal{O}(N\log_2 N)$ & $\mathcal{O}(N\log_2 N)$ \\
& T-count $\mathcal{O}(T)$ & -- & -- & $\mathcal{O}\left(N\log_2 N+N\log_2(1/\varepsilon)\right)$ \\
& Measurement Complexity & -- & -- & $O\left( \frac{\log_2(1/\delta)}{e^2} \right)$ \\
\midrule
\multirow{4}{*}{V{\small 2}M} 
& Qubits & -- & $2\log_2 N+1$ & $2\log_2 N$ \\
& Gates $\mathcal{O}g$ & -- & $\mathcal{O}(N^2\log_2 N)$ & $\mathcal{O}(N^2\log_2 N)$ \\
& T-count $\mathcal{O}(T)$ & -- & -- & $\mathcal{O}\left(N^2\log_2 N+N^2\log_2(1/\varepsilon)\right)$ \\
& Measurement Complexity & -- & -- & $O\left( \frac{\log_2 N+\log_2(1/\delta)}{e^2} \right)$ \\
\midrule
\multirow{4}{*}{M{\small 2}M} 
& Qubits & -- & $(N+2)\log_2 N+N$ & $3\log_2 N$ \\
& Gates $\mathcal{O}g$ & -- & $\mathcal{O}(N^3\log_2 N)$ & $\mathcal{O}(N^2\log_2 N)$ \\
& T-count $\mathcal{O}(T)$ & -- & -- & $\mathcal{O}\left(N^2\log_2 N+N^2\log_2(1/\varepsilon)\right)$ \\
& Measurement Complexity & -- & -- & $O\left( \frac{\log_2 N^2+\log_2(1/\delta)}{e^2} \right)$ \\
\midrule
\multirow{4}{*}{O{\small 2}M} 
& Qubits & -- & -- & $2\log_2 N+\log_2 K$ \\
& Gates $\mathcal{O}g$ & -- & -- & $\mathcal{O}\left(N^2\log_2 N+\sum_{k=0}^{K-1}G_{\mathrm{controlled}\text{-}B_k}\right)$ \\
& T-count $\mathcal{O}(T)$ & -- & -- & $\mathcal{O}\left(N^2\log_2 N+N^2\log_2(1/\varepsilon)+\sum_{k=0}^{K-1}T_{\mathrm{controlled}\text{-}B_k}\right)$ \\
& Measurement Complexity & -- & -- & $O\left( \frac{\log_2 (KN^2)+\log_2(1/\delta)}{e^2} \right)$ \\
\midrule
\multirow{4}{*}{SMM} 
& Qubits & -- & -- & $3\log_2 N+\log_2 L$ \\
& Gates $\mathcal{O}g$ & -- & -- & $\mathcal{O}(LN^2\log_2 N)$ \\
& T-count $\mathcal{O}(T)$ & -- & -- & $\mathcal{O}\left(LN^2\log_2 N+LN^2\log_2(1/\varepsilon)\right)$ \\
& Measurement Complexity & -- & -- & $O\left( \frac{\log_2 N^2+\log_2(1/\delta)}{e^2} \right)$ \\
\bottomrule
\end{tabular}
} %
\end{table*}

\section{Results}
\subsection{Overview of the QKMM algorithm and the family of quantum linear algebra operators}

In Fig.\ref{fig:fig1_framework}a, we compare the asymptotic complexity of the QKMM algorithm against classical matrix multiplication. It is important to note that classical algorithms typically count floating-point multiplications as their fundamental operation, whereas quantum algorithms employ elementary quantum gates as the basic computational unit. With this distinction in mind, we analyze the scaling behavior of these respective primitive operations with respect to the matrix dimension N. Notably, when constant factors are neglected, the gate complexity of QKMM—$O(N^2log_2N)$—surpasses that of the optimal classical matrix multiplication algorithm $O(N^{2.371339})$, with the crossover occurring at \(N\sim10^3\). At this scale, the output of QKMM is encoded in a quantum state; retrieving the embedded data requires additional measurements. In our amplitude encoding scheme, phase information is irrelevant to the readout process, thereby circumventing the prohibitive overhead associated with full quantum state tomography. Instead, the measurement task reduces to a classical sampling problem. Using mathematical tools such as Hoeffding’s inequality, we derive a measurement complexity of $O\left( \frac{1}{e^2}(\log_2 N^2+\log_2(1/\delta)) \right)$, with a rigorous proof provided in the Appendix and extended discussions available in ref. \cite{hoeffding1963probability,aaronson2007learnability}. Consequently, as illustrated in Fig.\ref{fig:fig1_framework}a, once N reaches sufficiently large scales (e.g., on the order of $10^9$), the total asymptotic complexity of QKMM—including measurement overhead—falls below that of the best-known classical matrix multiplication algorithms.

Fig.\ref{fig:fig1_framework}b illustrates the family of quantum linear algebra operators developed around the QKMM (also referred to as M{\small 2}M) framework. Within this family, V{\small 2}V and V{\small 2}M serve as the basic operators, while M{\small 2}M forms the core of quantum matrix operations. Building on M{\small 2}M, we construct two advanced operators: one-to-many (O{\small 2}M) mapping and sequential matrix multiplication (SMM). Notably, O{\small 2}M underpins the core computation of multi-head attention mechanisms, and SMM provides the essential operational layer for deep neural networks. Table \ref{tab:resource_scaling} summarizes the resource overheads—specifically qubit complexity, gate complexity, T-gate complexity, and measurement complexity—associated with implementing these operators. Our approach demonstrates distinct advantages over established quantum algorithms such as the Swap Test and Hadamard Test, not only by significantly reducing qubit and gate consumption but also by realizing O{\small 2}M and SMM functionalities that are unattainable with these conventional methods. Fig.\ref{fig:fig1_framework}c schematically outlines the implementation logic of QKMM; complete quantum circuits for all operators are provided in the Methods section.

\subsection{Noiseless simulation: Performance baseline and resource-accuracy trade-off}

In noiseless ideal simulations, we established the performance baseline of the quantum-matrix-multiplication-driven operator hierarchy. Quantum computation involves two distinct resource components: the coherent resources required for circuit execution and the measurement cost required to extract classical information from quantum states. The former determines circuit complexity, whereas the latter governs output accuracy under finite-shot measurement. All inputs were normalized and loaded through amplitude encoding. Non-power-of-two dimensions were padded with zeros to \(2^m\) and \(2^n\) dimensions, where \(m,n\in\mathbb{Z}\), to match the register size. For each operator, 100 randomly generated instances were evaluated. Under noiseless conditions, quantum-state evolution remains ideal, and the observed measurement errors mainly originate from statistical estimation with finite shots.

We first quantified the resource requirements of a single coherent evolution. Based on the complexity derivations in the Supplementary Information, Table \ref{tab:resource_scaling} summarizes the qubit overhead, elementary gate complexity and fault-tolerant T-count of V{\small 2}V, V{\small 2}M, M{\small 2}M, O{\small 2}M and SMM. The elementary gate complexity characterizes the execution scale after circuit decomposition, whereas the T-count estimates the non-Clifford overhead required to implement rotations and controlled encoding operations within a Clifford+T fault-tolerant architecture. Across different operators, the resource requirements are determined by three structural factors: the number of controlled amplitude-encoding operations, the size of the index registers and the branch or cascade control structure. This analysis establishes the circuit-resource boundaries for subsequent accuracy evaluations.

After characterizing circuit resources, we investigated the impact of finite measurement on computational accuracy. The comparison between quantum amplitude extraction and the classical ground truth in Fig.\ref{fig:fig2_Noiseless}g shows agreement approaching machine precision, indicating that the QKMM circuit evolution itself introduces negligible numerical deviation under noiseless conditions. Therefore, the error observed in probability-based readout is primarily governed by finite-shot statistical measurement.

As the most fundamental output task, V{\small 2}V only requires estimating the computational-basis probability associated with a single inner product. Fig.\ref{fig:fig2_Noiseless}a demonstrates that, with 10,000 shots, the median error of V{\small 2}V remains at the \(10^{-2}\) level as the input dimension increases from \(N=1280\) to \(4096\), while maintaining a more concentrated error distribution than the Swap Test and Hadamard Test. This stability originates from the direct mapping of the target inner product to a designated probability branch, avoiding auxiliary-qubit interference measurements and indirect reconstruction procedures.

We next moved from vector outputs to matrix-level transformations to characterize the scaling of measurement overhead with problem size. V{\small 2}M requires the simultaneous estimation of \(N\) row-wise inner products, whereas M{\small 2}M resolves \(N^2\) output branches defined by combined row and column indices. Fig.\ref{fig:fig2_Noiseless}b reveals that the error of V{\small 2}M increases gradually with dimension while remaining consistently below that of the Hadamard Test. In contrast, the output space of M{\small 2}M expands quadratically with matrix dimension, reducing the effective number of samples allocated to each output branch. Consequently, its error increases more rapidly with dimension, reaching the \(10^{-1}\) level at \(N=64\), as demonstrated in Fig.\ref{fig:fig2_Noiseless}c. The continuous error reduction with increasing shots, demonstrated in Fig.\ref{fig:fig2_Noiseless}e, further confirms that the observed degradation is primarily caused by finite-shot measurement uncertainty rather than circuit evolution errors. It should be noted that target inner products and matrix elements remain fully encoded in quantum amplitudes during coherent evolution. The probability readout adopted here directly estimates the squared amplitude magnitude, while complete amplitude reconstruction would require quantum-state tomography and introduce additional measurement overhead. Therefore, the measurement analysis presented here characterizes the cost of probability-based classical readout rather than the intrinsic information content preserved within the quantum state.

Beyond single matrix transformations, we further evaluated O{\small 2}M and SMM to examine scalability along parallel branching and sequential cascading dimensions. Fig.\ref{fig:fig2_Noiseless}d highlights that the error of O{\small 2}M increases with the number of parallel matrices, reflecting the redistribution of finite measurement resources among multiple output branches. A similar trend is observed for SMM in Fig.\ref{fig:fig2_Noiseless}f, where the accumulated error increases with cascade depth \(L\) and becomes more pronounced for higher-dimensional matrices. These results indicate that coherent matrix propagation can be maintained without intermediate measurements, while final readout accuracy remains constrained by output dimensionality, cascade depth and finite-shot statistics.

The runtime analysis provided in the Supplementary Information reveals consistent scaling trends. V{\small 2}V and V{\small 2}M maintain relatively low simulation costs as the dimension increases, whereas the average time per matrix operation in O{\small 2}M and SMM gradually decreases and saturates with increasing branch number or cascade depth. This behaviour indicates that shared encoding, index scheduling and circuit reuse can reduce repeated preparation overhead. However, these measurements represent computational costs in a classical simulation environment rather than execution times on quantum hardware. As the number of qubits increases, high-dimensional quantum-state evolution remains the dominant limitation for numerical simulation scalability.

Overall, noiseless simulations demonstrate that the proposed operator hierarchy reduces redundant circuit construction through coherent parallelism and resource reuse, while also revealing the measurement limitations associated with explicit reconstruction of complete classical outputs. In end-to-end quantum neural-network inference, intermediate states can instead remain coherently propagated without layer-wise measurement and re-encoding. Therefore, the measurement overhead discussed above primarily applies to full classical output reconstruction rather than task-level quantum inference. The following section further evaluates the application of this operator hierarchy to neural-network inference.

\begin{figure*} 
\centering
\includegraphics[width=1.0\textwidth]{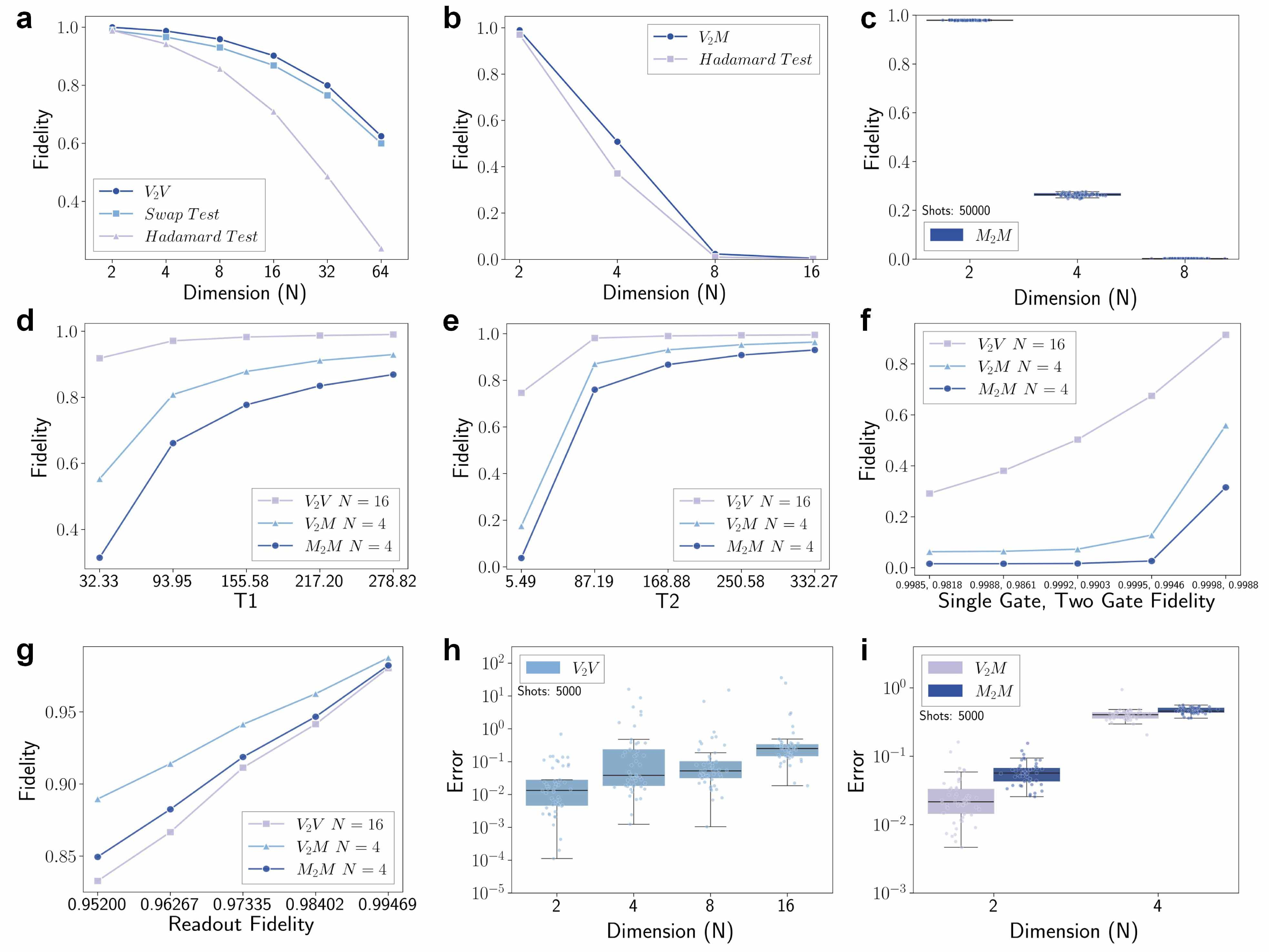} 
	\caption{\textbf{Performance evaluation of the QKMM operator system under noisy simulation and execution on the Origin Wukong quantum processor.} \textbf{a–c}, Fidelity with increasing input dimension \(N\), showing the robustness advantage of V{\small 2}V over the Swap Test and Hadamard Test (\textbf{a}), V{\small 2}M over the Hadamard Test (\textbf{b}). \textbf{d–g}, Impact of physical noise parameters on fidelity, including relaxation time $T_1$ (\textbf{d}), dephasing time $T_2$ (\textbf{e}), single- and two-qubit gate fidelities (\textbf{f}), and readout fidelity (\textbf{g}). \textbf{h,i}, Experimental error distributions evaluated on the Origin Wukong 180-qubit superconducting quantum hardware for V{\small 2}V (\textbf{h}), as well as V{\small 2}M and M{\small 2}M (\textbf{i}) across different dimensions $N$.}
	\label{fig:fig3_Noisy} 
\end{figure*}

\subsection{Noisy simulation: Characterization of robustness under NISQ-era constraints}

After establishing the ideal resource baseline, we further evaluated the noise robustness of the QKMM operator hierarchy under NISQ conditions. Since noisy simulations must simultaneously capture quantum-state evolution and noise-channel effects, their classical computational cost increases rapidly with circuit size. We therefore selected three representative operators, V{\small 2}V, V{\small 2}M and M{\small 2}M, for systematic evaluation. The noise model was constructed from calibration parameters of the IBM Heron superconducting quantum processor, incorporating \(T_1\) relaxation, \(T_2\) decoherence, single- and two-qubit gate fidelities and readout fidelity, with details provided in the Supplementary Information. To distinguish noise-induced state degradation from finite-shot statistical fluctuations, quantum-state fidelity was used as the primary metric, complemented by the measurement-error and runtime analyses presented in the Supplementary Information.

Under identical noise conditions, we first examined the dimensional scalability of different operators. Fig.\ref{fig:fig3_Noisy}a shows that the fidelity of V{\small 2}V gradually decreases with increasing dimension, while remaining consistently higher than that of the Swap Test and Hadamard Test. This advantage originates from its direct probability-mapping structure: V{\small 2}V avoids auxiliary qubits and additional interference-based estimation, resulting in shallower circuits and fewer controlled operations, which reduces accumulated noise during execution.

When the computation is extended to matrix-level mappings, noise sensitivity becomes substantially stronger. Fig.\ref{fig:fig3_Noisy}b reveals that the fidelity of V{\small 2}M decreases noticeably with increasing dimension, indicating that additional output branches introduce increased controlled encoding and circuit-depth overhead. M{\small 2}M further amplifies this effect. As shown in Fig.\ref{fig:fig3_Noisy}c, its fidelity rapidly deteriorates as the dimension increases and approaches failure at \(N=8\). This rapid degradation indicates that, under current NISQ noise levels, the executable scale of matrix multiplication is strongly constrained. The dominant limitations arise from increasing circuit depth, a larger number of controlled encoding operations and accumulated two-qubit gate errors.
To identify the contributions of different noise sources, we performed a parameter-sensitivity analysis. Fig.\ref{fig:fig3_Noisy}d,e show that increasing \(T_1\) and \(T_2\) improves the fidelity of all three operators, with matrix-level operators exhibiting a stronger response. This indicates that their performance is more dependent on coherence time. Fig.\ref{fig:fig3_Noisy}f demonstrates that reducing single- and two-qubit gate fidelities substantially decreases the output fidelity of V{\small 2}M and M{\small 2}M, whereas V{\small 2}V is affected to a lesser extent. This further identifies two-qubit gate count and circuit depth as key factors limiting matrix-level scalability. Fig.\ref{fig:fig3_Noisy}g shows that improving readout fidelity continuously enhances the measurement fidelity of all three operators. When readout fidelity approaches \(99.5\%\), the output fidelity exceeds 0.98, whereas lower readout fidelity further amplifies errors in multi-branch probability resolution.

The relative-error and runtime analyses in the Supplementary Information further support these observations. V{\small 2}V exhibits the highest stability in both error growth and simulation-time scaling, V{\small 2}M maintains moderate scalability, whereas M{\small 2}M simultaneously suffers from stronger noise sensitivity and higher classical simulation costs as the dimension increases. These results indicate that scaling matrix-level quantum computation is constrained by both physical noise and the computational cost of classical validation.

Overall, NISQ noise does not alter the theoretical complexity scaling of the QKMM operator hierarchy but significantly restricts the scale of reliable execution. The operators exhibit a clear robustness hierarchy: V{\small 2}V is the least sensitive to noise, V{\small 2}M is increasingly affected by controlled mappings, and M{\small 2}M is the most vulnerable due to its larger number of index-controlled operations and two-qubit gates. Further scaling of matrix-level quantum computation will require simultaneous improvements in two-qubit gate fidelity, coherence time, circuit compilation efficiency and error-mitigation techniques.

\subsection{Verification on Superconducting Quantum Computer}

After completing the ideal and noisy simulations, we further executed the V{\small 2}V, V{\small 2}M and M{\small 2}M circuits on the 180-qubit Origin Wukong superconducting quantum computer to assess the physical executability of the operator hierarchy under realistic noise, finite coherence times, hardware-connectivity constraints and gate-calibration errors. The circuits were compiled into the native $R_{\phi}$ and CZ gate set and preferentially mapped onto connected physical-qubit subgraphs with favourable calibration characteristics to minimize additional routing overhead. The physical qubit connectivity topology and detailed calibration parameters are provided in the Supplementary Information. To reduce dependence on specific input instances, 50 independently generated and normalized inputs were evaluated for each operator at every tested dimension, with 5,000 shots used for all measurements.

As shown in Fig.\ref{fig:fig3_Noisy}h, V{\small 2}V maintains a relatively low error level across the tested dimensions. At \(N=2\), the median relative error remains on the order of \(10^{-2}\). As the dimension increases, the error distribution shifts upward and reaches the \(10^{-1}\) level at \(N=16\), accompanied by increased variation across input instances. These results indicate that the direct inner-product readout of V{\small 2}V retains good numerical stability on real hardware, although increasing encoding scale and circuit resources progressively expose the calculation to gate errors, decoherence and readout noise.

As the computation extends from inner products to matrix-level mappings, hardware-induced degradation becomes substantially more pronounced. Fig.\ref{fig:fig3_Noisy}i shows that V{\small 2}M exhibits consistently higher errors than V{\small 2}V under the same dimensional and measurement conditions, while M{\small 2}M shows a further increase in error. For both matrix-level operators, the error grows markedly with input dimension. This hierarchy is consistent with their circuit structures: V{\small 2}M and M{\small 2}M require additional index registers, more controlled amplitude-encoding operations and a larger number of two-qubit gates, while M{\small 2}M must additionally resolve multiple output branches defined jointly by row and column indices. These structural requirements make matrix-level operators more susceptible to finite \(T_2\), accumulated two-qubit gate errors, topology-induced routing overhead and multi-qubit readout noise.

\begin{figure*} 
\centering
\includegraphics[width=1.0\textwidth]{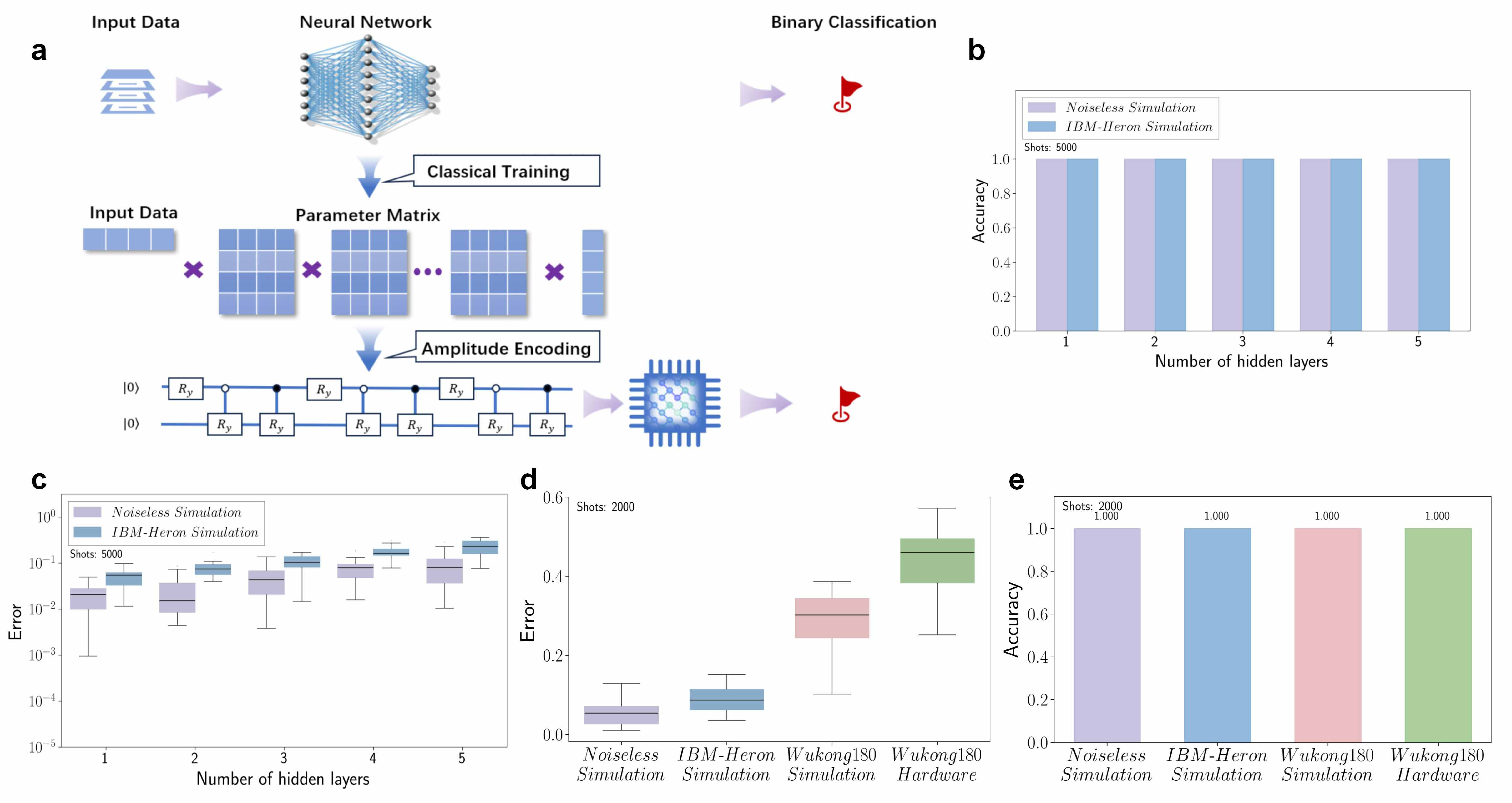} 
	\caption{\textbf{QKMM-driven quantum neural-network inference under noisy simulation and superconducting hardware.} \textbf{a}, Hybrid computational framework embedding the QKMM operator system into neural network inference. \textbf{b,c}, Classification accuracy (\textbf{b}) and output error distribution (\textbf{c}) as functions of hidden layer depth $H$ under noiseless and IBM-Heron noisy simulations. \textbf{d,e}, Output error distribution (\textbf{d}) and classification accuracy (\textbf{e}) for a two-hidden-layer network across four backends: noiseless simulation, IBM Heron noise model, Origin Wukong noise model, and the Origin Wukong 180-qubit superconducting quantum processor.}
	\label{fig:fig4_iris} 
\end{figure*}

Overall, the results obtained on the superconducting quantum processor reveal the same robustness hierarchy observed in the noisy simulations: V{\small 2}V is the most stable, followed by V{\small 2}M, whereas M{\small 2}M is the most sensitive to hardware noise. These results demonstrate that current quantum hardware is capable of executing fundamental quantum operators. However, as the computational structure extends from vector-level operations to matrix-level transformations, the increased number of controlled operations and circuit depth impose more stringent requirements on two-qubit gate fidelity, coherence time and readout accuracy.

\subsection{Application to Neural Network Inference}

The preceding noiseless analysis shows that explicit reconstruction of complete vectors or matrices remains limited by finite-shot measurement. In an end-to-end computational pipeline, however, intermediate results can remain coherently encoded in the quantum state and be directly propagated into subsequent operations without layer-wise measurement and re-encoding. Leveraging this coherent propagation, we further embed the quantum-matrix-multiplication-driven operator hierarchy into a neural-network inference pipeline to assess its ability to extend from isolated linear operations to an end-to-end computational task.

As shown in Fig.\ref{fig:fig4_iris}a, we constructed a hybrid classical-training–quantum-inference framework. The model parameters are first optimized on a classical computer and then fixed and encoded into the quantum circuit. The quantum inference circuit follows a V{\small 2}M–SMM–M{\small 2}V architecture: V{\small 2}M performs the initial feature mapping, SMM propagates the intermediate features across hidden layers through sequential matrix multiplications, and M{\small 2}V serves as the reverse mapping of V{\small 2}M, projecting the final feature representation into the classification output. To isolate the contribution of quantum matrix multiplication to coherent linear propagation, the neural-network architecture used in this experiment does not include nonlinear activation functions. Throughout the inference process, intermediate quantum states are propagated coherently and measured only at the final classification stage, thereby avoiding layer-by-layer reconstruction of intermediate vectors or matrices and the associated repeated measurement and re-encoding overhead.

Building on this framework, we constructed a two-feature binary classification task using the Iris dataset to examine the effects of network depth and hardware noise on end-to-end quantum inference. All experiments used the same data preprocessing, model parameters and classification criterion. To quantify the effect of network depth, we used the number of sequential matrix transformations in SMM to represent the hidden-layer depth and evaluated its impact on coherent feature propagation. Fig.\ref{fig:fig4_iris}b shows that, with 5,000 shots, both the noiseless simulation and the IBM-Heron noisy simulation maintain a classification accuracy of $100\%$ across all tested depths. Fig.\ref{fig:fig4_iris}c, however, shows that the output error generally increases with the number of hidden layers, with the noisy results consistently exceeding their noiseless counterparts. This trend indicates that errors arising from the increasing number of controlled encoding operations and two-qubit gates, together with longer circuit execution times, accumulate as the matrix chain deepens, although they remain insufficient to alter the final predicted labels. Thus, degradation in numerical precision emerges before a measurable decline in task-level classification accuracy.

To further compare the impact of different execution environments on end-to-end inference, we fixed the network to two hidden layers and used a V{\small 2}M–M{\small 2}M–M{\small 2}V operator chain. The same task was evaluated under noiseless simulation, the IBM-Heron noise model, an Origin Wukong 180 parameter-matched noise model, and the physical Origin Wukong 180 superconducting processor, with 2,000 shots per sample. Fig.\ref{fig:fig4_iris}d shows a progressive increase in output error from noiseless simulation to real-hardware execution. On the physical processor, the median error reaches approximately 0.46 and the distribution broadens substantially across samples. Nevertheless, Fig.\ref{fig:fig4_iris}e shows that the classification accuracy remains $100\%$ under all four execution conditions. These results indicate that the classification decision is tolerant to a certain level of numerical perturbation in this task, while also demonstrating that stable class labels do not necessarily imply high-precision numerical outputs.

Overall, these experiments verify that the V{\small 2}M–SMM–M{\small 2}V operator chain can perform end-to-end inference without layer-wise measurement of intermediate results. As the number of hidden layers increases, the dominant limitation shifts from finite-shot measurement toward the accumulation of errors associated with greater circuit depth and hardware noise. Evaluating both classification accuracy and numerical output error is therefore essential for characterizing the practical reliability of quantum inference under current NISQ conditions.

\section{Discussion}

Quantum acceleration of matrix multiplication extends beyond reducing the complexity of individual operations; it provides a foundation for scalable quantum linear-algebra frameworks in which quantum transformations can be composed, propagated and reused across complex computational workflows. QKMM provides a framework toward this goal by representing matrix transformations as composable quantum operators, allowing linear transformations at different levels to be continuously executed within a unified quantum architecture. Measured by the number of elementary quantum gates, QKMM achieves a coherent circuit complexity of $O(N^2\log_2N)$, exhibiting a lower asymptotic scaling than the current best-known classical matrix multiplication algorithm with complexity $O(N^{2.371339})$. Built upon this framework, the quantum linear-algebra operator hierarchy supports vector operations, matrix transformations and sequential propagation, and demonstrates the possibility of coherently propagating intermediate features in neural-network inference without layer-wise measurement and re-encoding. This offers an alternative paradigm to the conventional “compute–measure–repreparation” workflow.

However, the transition from theoretical resource advantages to practical computational benefits remains challenging. The complexity advantage measured by quantum gate count reflects an improvement in algorithmic resource scaling rather than a direct runtime advantage on current quantum processors. Unlike quantum algorithms that rely on quantum random access memory (QRAM) to enable large-scale element-wise random access, QKMM does not require QRAM as an essential assumption of the computational model. Instead, it performs matrix transformations through global quantum-state encoding and coherent quantum circuits, thereby avoiding the additional resource requirements associated with introducing a QRAM oracle. Nevertheless, quantum state preparation, data encoding, measurement overhead, finite coherence time, gate errors, and hardware connectivity constraints continue to limit the practical scalability of large-scale quantum matrix computation. For tasks requiring explicit reconstruction of complete classical matrices, classical methods may still achieve lower output reconstruction costs; however, for quantum subroutines requiring only matrix–vector multiplication or inner-product estimation, QKMM provides reduced coherent computational overhead. Therefore, the value of QKMM does not lie in replacing all classical matrix operations, but rather in providing a coherent computational pathway distinct from classical data exchange paradigms, where quantum information can be continuously preserved and directly utilized in subsequent computations.

Furthermore, matrix multiplication represents one of the most fundamental compute-intensive operations underlying modern artificial intelligence and scientific computing. The significance of quantum matrix computation extends beyond accelerating individual algorithms, providing a basis for exploring new approaches to addressing broader computational bottlenecks with quantum resources. We refer to this emerging direction as “quantum-intensive computing”, where quantum computing resources are leveraged to explore acceleration for large-scale, computation-dominated tasks, a field that remains at an early stage of development. As the demands of large-scale artificial intelligence model training and scientific computing continue to grow, demonstrating verifiable quantum advantages in compute-intensive tasks will be an important step toward advancing quantum computing from specialized algorithmic acceleration to a more general computational paradigm.


\begin{figure*}
\centering
\subfigure{
\includegraphics[width=0.85\textwidth]{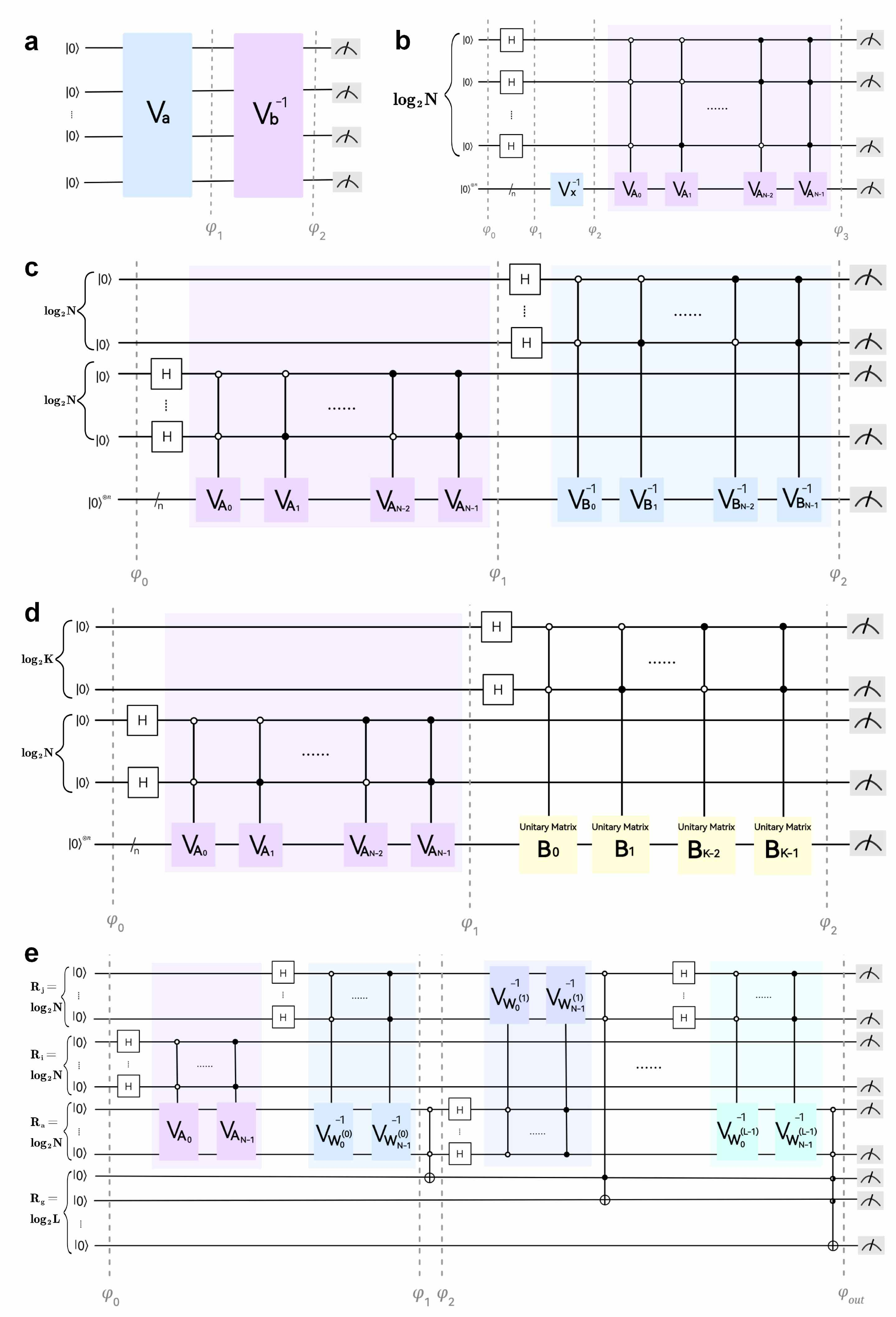}}
\caption{\textbf{Quantum Circuits for the Foundational Operator System of Quantum Neural Networks.} \textbf{a}, The quantum circuit for V{\small 2}V. \textbf{b}, The quantum circuit for V{\small 2}M. \textbf{c}, The quantum circuit for M{\small 2}M. \textbf{d}, The quantum circuit for O{\small 2}M. \textbf{e}, The quantum circuit for SMM.}
\label{fig:cir}
\end{figure*}

\section{Methods}
\subsection{Vector to Vector (V{\small 2}V)}
As the fundamental building block of the network's operator framework, the Vector to Vector (V{\small 2}V) operator evaluates the inner product modulus of two high-dimensional vectors to implement quantum-kernel-based neuronal functionality. This core architecture is based on the quantum kernel-based inner product modulus method \cite{havlivcek2019supervised}.

Given any two normalized vectors $|\mathbf{a}\rangle ,| \mathbf{b}\rangle \subseteq \mathbb{R}^N$, amplitude encoding is used to map these normalized vectors into quantum basis states $|0\rangle ^{\otimes n}$. The encoded quantum states can be expressed as:
\begin{equation}
|\mathbf{a}\rangle = V_a|0\rangle ^{\otimes n}, \quad 
|\mathbf{b}\rangle= V_b|0\rangle ^{\otimes n}
\end{equation}
where \(n=\lceil \log_2 N \rceil \), \( V_a \) and \( V_b \) are the unitary transformations corresponding to the amplitude encoding of \( \mathbf{a} \) and \( \mathbf{b} \), respectively. The quantum circuit for this process is illustrated in Fig.\ref{fig:cir}a, and the evolution of the circuit can be described as follows:
\begin{gather}  
|\varphi _1 \rangle =V_a|0\rangle ^{\otimes n}\\
|\varphi _2 \rangle ={V_b}^{-1}|\varphi _1 \rangle={V_b}^{-1}V_a|0\rangle ^{\otimes n}
\end{gather} 
then the inner product is given by:
\begin{equation}
\begin{split}
\left< \mathbf{a} | \mathbf{b}  \right> 
&= \left< \mathbf{a} |{V_b}{V_b}^{-1} | \mathbf{b}  \right> =\left< \mathbf{a} |{V_b}|0\right> ^{\otimes n}\\
&=\left< \varphi _2 \right. \left| {V_b}^{\dagger}V_b\left| \left. 0 \right> ^{\otimes n} \right. \right. =\left< \varphi _2 \right. \left| \left. 0 \right> ^{\otimes n} \right. =c_0
\end{split}
\label{eq-inner_product}
\end{equation}
Here, \( c_0 \) represents the coefficient of the superposition state \( |\varphi _2 \rangle \) in the ground state \( |0\rangle ^{\otimes n} \). Therefore, when we perform a measurement on the entire circuit, the obtained probability is \( |c_0|^2 \).

\subsection{Vector to Matrix (V{\small 2}M)}
By extending the V{\small 2}V architecture into a high-dimensional topology for matrix–vector multiplications, the Vector to Matrix (V{\small 2}M) operator implements the fully quantum forward feature mapping of an individual input sample through the hidden-layer space.

For a matrix $A \in \mathbb{R}^{N \times N}$ and a vector $\mathbf{x} \in \mathbb{R}^N$, their linear transformation $\mathbf{b} = A\mathbf{x}$ is evaluated element-wise as $b_i = \langle A_{i\cdot} | \mathbf{x} \rangle$, where $A_{i\cdot}$ represents the $i$-th row of $A$.

The V{\small 2}M operator fundamentally relies on row-wise decomposition of the weight matrix, utilizing index registers and multi-controlled networks for fully parallel quantum storage within a single circuit. Subsequently, upon applying the inverse amplitude encoding of the input vector, quantum kernel methods are employed to concurrently compute the inner products between the matrix rows and the vector, mapping these results onto the ground-state amplitudes of a joint superposition state. Requiring only $\lceil \log_2 N \rceil$ ancilla qubits, this architecture bypasses traditional circuit-stacking limitations to efficiently implement single-sample forward linear mapping. The quantum circuit is shown in Fig.\ref{fig:cir}b, evolving as follows:
\begin{gather}  
\left| \left. \varphi _0 \right> = \right. \left| \left. 0 \right> \right. ^{\otimes n}\left| \left. 0 \right> \right. ^{\otimes n}
\\
\left. |\varphi _1 \right> =\left( \frac{1}{2} \right) ^{\frac{n}{2}}\sum_{i=0}^{N-1}{\left| \left. i \right> \right.}\left| \left. 0 \right> \right. ^{\otimes n}
\\
\left. |\varphi _2 \right> =\left( \frac{1}{2} \right) ^{\frac{n}{2}}\sum_{i=0}^{N-1}{\left| \left. i \right> \otimes V^{-1}_x \right.}\left| \left. 0 \right> \right. ^{\otimes n}
\\
\left. |\varphi _3 \right> =\left( \frac{1}{2} \right) ^{\frac{n}{2}}\sum_{i=0}^{N-1}{\left| \left. i \right> \otimes \right.}V_{A_i}V^{-1}_x\left| \left. 0 \right> \right. ^{\otimes n}
\end{gather} 
where $n = \lceil \log_2 N \rceil$, and the encoding unitaries $V_x, V_{A_{i\cdot}}$ satisfy $|x\rangle = V_x |0\rangle^{\otimes n}$ and $|A_{i\cdot}\rangle = V_{A_{i\cdot}} |0\rangle^{\otimes n}$. Following the same logic as in Eq. \ref{eq-inner_product}, the inner product \(\left< A_{i\cdot} \right. \left| \left. x \right> \right. \) is stored in the amplitude of the superposition state with respect to the ground state \( |i\rangle|0\rangle ^{\otimes n} \). Therefore, the probability of observing the ground state $\left| \left. i \right> \right. \left| \left. 0 \right> \right. ^{\otimes n}$ corresponds to the $\frac{1}{N}|\left< A_{i\cdot} \right. \left| x \right> |^2$.

\subsection{Matrix to Matrix (M{\small 2}M)}
To accommodate the modern machine learning paradigm heavily reliant on batch training, the Matrix to Matrix (M{\small 2}M) operator concurrently computes both input samples and network weights within a single quantum circuit. 

The product $C=AB$ of two matrices $A, B \in \mathbb{R}^{N \times N}$ is defined element-wise by $c_{ij} = \langle A_{i\cdot} | B_{\cdot j} \rangle$, where $A_{i\cdot}$ and $B_{\cdot j}$ represent the $i$-th row of $A$ and the $j$-th column of $B$.

By employing dual index registers to encode the rows of matrix $A$ and columns of matrix $B$, the M{\small 2}M operator seamlessly integrates their full structural information into a single circuit via multi-controlled amplitude encoding. This approach eliminates the circuit-stacking and encoding redundancies of traditional algorithms, concurrently computing the cross inner products of all row and column vectors during a single circuit evolution. Compressing the ancilla qubit overhead to only $2 \log_2 N$, this design delivers fully parallel matrix multiplication at an ultra-low hardware cost, constituting the core acceleration engine for large-scale batch processing within quantum neural networks. The quantum circuit is illustrated in Fig.\ref{fig:cir}c, and its specific evolution process is as follows:
\begin{footnotesize}
\begin{gather}  
\left| \left. \varphi _0 \right> = \right. \left| \left. 0 \right> \right. ^{\otimes n}\left| \left. 0 \right> \right. ^{\otimes n}\left| \left. 0 \right> \right. ^{\otimes n}
\\
\left. |\varphi _1 \right> =\left( \frac{1}{2} \right) ^{\frac{n}{2}}\left| \left. 0 \right> \right. ^{\otimes n}\otimes \sum_{i=0}^{N-1}{\left| \left. i \right> \otimes V_{A_i} \right.}\left| \left. 0 \right> \right. ^{\otimes n}
\\
\begin{split}
\left. |\varphi _2 \right> &=\left( \frac{1}{2} \right) ^{n}\sum_{j=0}^{N-1}{\left| \left. j \right> \otimes \right.}\sum_{i=0}^{N-1}{\left| \left. i \right> \otimes V_{B_j}^{-1}V_{A_i} \right.}\left| \left. 0 \right> \right. ^{\otimes n}
\\
&=\left( \frac{1}{2} \right) ^{n}\sum_{j=0}^{N-1}{\sum_{i=0}^{N-1}{\left| \left. j \right> \right. \left| \left. i \right> \right.}}V_{B_j}^{-1}V_{A_i}\left| \left. 0 \right> \right. ^{\otimes n}
\end{split}
\end{gather} 
\end{footnotesize}
here, \( n=\lceil \log_2 N \rceil  \), \( V_{B_j} \) and \( V_{A_i} \) represent the unitary matrices for data encoding, with \( \left| \left. A_{i\cdot} \right> =V_{A_i} \right. \left| \left. 0 \right> \right. ^{\otimes n} \) and \( \left| \left. B_{\cdot j} \right> =V_{B_j} \right. \left| \left. 0 \right> \right. ^{\otimes n} \). The inner product \( \left< A_{i\cdot} \right. \left| \left. B_{\cdot j} \right> \right.  \) is stored in the amplitude \( \left. |\varphi _2 \right>  \) of the superposition state with respect to the ground state \( |j\rangle|i\rangle|0\rangle^{\otimes n} \). Therefore, the final result can be obtained through repeated measurements of the circuit.

\subsection{One-to-many matrix multiplication (O{\small 2}M)}

The One-to-many matrix multiplication (O{\small 2}M) operator enables a row-normalized matrix to concurrently interact with a set of distinct unitary matrices within a unified quantum circuit. This architecture exhibits a natural topological mapping to both the multi-head attention blocks of Transformer networks and parallel multi-task quantum neural networks.

Let \( A \) be an \( N \times N \) matrix, and \(\{ B_0, B_1, \cdots , B_{K-1} \}\) be a set of $K$ independent $N \times N$ unitary matrices representing distinct attention heads or task weights. Their product \( C_i \) is defined as:
\begin{gather}  
C_i=A\cdot B_i\,\,\,\,\,\,\,\,\,\,(i=0,1,\cdots ,K-1)
\end{gather} 
The circuit execution decomposes the operation into the product of the $j$-th row of matrix $A$ and the unitary matrix $B_k$. The normalized vector $A_{j}$ is encoded into the ground state $|0\rangle^{\otimes n}$, expressed as $|A_{j}\rangle=V_{A_j}|0\rangle^{\otimes n}$. Here 
$n=\lceil \log_2 N \rceil$, $V_{A_j}$ is the unitary matrix for amplitude encoding. The evolution of the circuit proceeds as follows:
\begin{gather}  
|\varphi _1 \rangle =V_{A_j}|0\rangle ^{\otimes n}\\
|\varphi _2 \rangle ={B_k}|\varphi _1 \rangle={B_k}|A_j \rangle
\end{gather} 
When measuring the quantum state \( |\varphi _2\rangle \), the probability of obtaining outcome \( |i\rangle \) is $\left| \left< B_{k_{i\cdot}} \right. \left| \left. A_j \right> \right. \right|^2$. Applying this principle to the M{\small 2}M algorithm enables parallel processing of multiple matrices. The quantum circuit is illustrated in Fig.\ref{fig:cir}d.

The O{\small 2}M operator bypasses redundant amplitude encoding processes by innovatively introducing a dedicated "head-index register" to directionally schedule the conditional execution of unitary operations $B_k$ within a fully coherent superposition state. Demonstrating exceptional resource scaling efficiency, this architecture requires a mere $\lceil \log_2 N \rceil + \lceil \log_2 K \rceil$ ancilla index qubits, $N$ multi-controlled amplitude encodings, and $K$ multi-controlled unitary matrix blocks. This enables the system to concurrently extract the complete attention map across all $K$ attention heads while preventing destructive collapse within individual subspace branches.

\subsection{Sequential-MM (SMM)}
As the top-level design of the QNN architecture, the Sequential-MM (SMM) operator executes the serial matrix chain multiplications required for deep multi-layer networks. To resolve the computational delays and severe decoherence triggered by the "inter-layer measurement and subsequent re-encoding" cycle in conventional hybrid networks, SMM utilizes a multi-stage controlled cascading topology. This design eliminates all intermediate measurements, requiring only $l=\lceil\log_2 L\rceil$ additional ancilla qubits over the M{\small 2}M operator to achieve lossless matrix chain multiplication under full coherence, thereby overcoming the fundamental physical bottleneck of cascaded forward propagation.

Given an input matrix $A \in \mathbb{R}^{N \times N}$ and hidden-layer weight matrices $\{W^{(0)}, W^{(1)}, \dots, W^{(L-1)}\}$, forward propagation aims to compute:
\begin{equation}
Y = W^{(L-1)} W^{(L-2)} \dots W^{(0)} A
\end{equation}
The quantum circuit is shown in Fig.\ref{fig:cir}e.

The system is initialized to the all-zero ground state $|\varphi_0\rangle = |0\rangle_a |0\rangle_i |0\rangle_j |0\rangle_g$ (where $n = \log_2 N$), consisting of a data register $R_a$ ($n$ qubits), a row-index register $R_i$ ($n$ qubits), a column-index register $R_j$ ($n$ qubits), and a Gray-code control register $R_g$ ($l$ qubits).

Reusing the M{\small 2}M operator output, the first layer evolves the system into the superposition state:
\begin{equation}
|\varphi_1\rangle = \frac{1}{N} \sum_{i,j} |i\rangle_i |j\rangle_j \left( V_{W_j^{(0)}}^\dagger V_{A_i} |0\rangle_a \right) |0\rangle_g    
\end{equation}
By the inner-product projection of amplitude encoding, when the data register is in state $|0\rangle_a$, its amplitude satisfies:
\begin{equation}
\langle0|_a V_{W_j^{(0)}}^\dagger V_{A_i} |0\rangle_a = \langle W_j^{(0)} | A_i \rangle = \big(W^{(0)}A\big)_{ji} 
\end{equation}
yielding the intermediate matrix $Z^{(0)} = W^{(0)}A$.

To seamlessly transfer preceding features and invoke $W^{(1)}$, the circuit employs a multi-controlled Gray code and a dynamic register-swapping mechanism. A non-destructive conditional projection is implemented via an operator strictly controlled by the data register's all-zero state $|0\rangle_a$. Swapping $R_a \longleftrightarrow R_j$ then smoothly maps the system state to:
\begin{equation}
|\varphi_2\rangle = \frac{1}{N} \sum_{i,j} |i\rangle_i |j\rangle_j |Z^{(0)}_{ji}\rangle_a |g_1\rangle_g
\end{equation}
During forward propagation, the algorithm chains forward mappings via alternating controlled amplitude encodings and Hadamard structures. After $L$ recursive expansions, the final fully coherent output state is:
\begin{footnotesize}
\begin{equation}
|\varphi_{out}\rangle = \frac{1}{N^{\frac{L+1}{2}}} \sum_{i,j} |i\rangle_i |j\rangle_j \left( \prod_{m=0}^{L-1} V_{W_j^{(m)}}^\dagger \right) V_{A_i} |0\rangle_a |g_{L-1}\rangle_g
\end{equation}
\end{footnotesize}
where the zero-state amplitude satisfies:
\begin{footnotesize}
\begin{equation}
\langle0|_a \left( \prod_{m=0}^{L-1} V_{W_j^{(m)}}^\dagger \right) V_{A_i} |0\rangle_a = \left[ W^{(L-1)} W^{(L-2)} \dots W^{(0)} A \right]_{ji}
\end{equation}    
\end{footnotesize}
This completes the fully coherent chain propagation of the deep feature matrix $Y$.

This closed-loop framework—integrating data-register ground-state gating, hierarchical subspace dynamic projection, and pointer-exchange reuse—fully circumvents intermediate-measurement-induced decoherence and classical re-encoding latencies, successfully preserving the optimal gate complexity of $O(LN^2 \log_2 N)$ for deep networks.

\section*{Acknowledgments}
Ding Liu is grateful to Shi-ju Ran for helpful discussions, and acknowledges the support from Origin Quantum Lab. This work was supported by Tianjin Natural Science Foundation of China (20JCYBJC00500) and the Science \& Technology Development Fund of Tianjin Education Commission for Higher Education (2018KJ217).

\bibliography{bibliography}



\end{document}